\documentclass[conference]{IEEEtran}
\IEEEoverridecommandlockouts
\usepackage{cite}
\usepackage{amsmath,amssymb,amsfonts}
\usepackage{algorithmic}
\usepackage{graphicx}
\usepackage{textcomp}
\usepackage{xcolor}
\def\BibTeX{{\rm B\kern-.05em{\sc i\kern-.025em b}\kern-.08em
    T\kern-.1667em\lower.7ex\hbox{E}\kern-.125emX}}

\renewcommand{\baselinestretch}{0.95}

\begin{document}

\title{Digital Twin for Real-Time Security Assessment and Flexibility Activation in the Bornholm Distribution System\\
\thanks{This work was supported by the European Union’s Horizon Europe programme under the SYNERGIES project (Grant Agreement No. 101069839) and by the European Union’s LCE Policy Support Programme under the ODEON project (Grant Agreement No. 101136128).}
}

\iffalse
\author{\IEEEauthorblockN{1\textsuperscript{st} Anosh Arshad Sundhu}
\IEEEauthorblockA{\textit{Department of Wind and Energy Systems} \\
\textit{Technical University of Denmark (DTU)}\\
Kgs. Lyngby, Denmark \\
anosu@dtu.dk}
\and
\IEEEauthorblockN{2\textsuperscript{nd} Ayseg{\"u}l Kahraman}
\IEEEauthorblockA{\textit{Department of Wind and Energy Systems} \\
\textit{Technical University of Denmark (DTU)}\\
Kgs. Lyngby, Denmark \\
ayska@dtu.dk}
\and
\IEEEauthorblockN{3\textsuperscript{rd} Spyros Chatzivasileiadis}
\IEEEauthorblockA{\textit{Department of Wind and Energy Systems} \\
\textit{Technical University of Denmark (DTU)}\\
Kgs. Lyngby, Denmark \\
spchatz@dtu.dk}
}
\fi

\author{
\IEEEauthorblockN{
Anosh Arshad Sundhu,
Ayseg{\"u}l Kahraman,
Spyros Chatzivasileiadis
}
\IEEEauthorblockA{
\textit{Department of Wind and Energy Systems}\\
\textit{Technical University of Denmark (DTU)}\\
Kgs. Lyngby, Denmark \\
\{anosu, ayska, spchatz\}@dtu.dk
}
}

\maketitle

\begin{abstract}
The increasing penetration of distributed energy resources (DERs) is transforming distribution networks into actively managed systems, introducing challenges related to voltage regulation, thermal loading limits, and operational security. This paper presents the development and implementation of a real-time Digital Twin (DT) for security assessment and coordinated flexibility activation in active distribution networks, demonstrated on the Bornholm Island system using real measurement data. The implemented DT integrates network topology and smart meter measurements to perform security assessment under normal operation and N-1 contingencies, and to determine corrective and preventive flexibility actions using an optimization-based approach. Results show that load variation and contingency scenarios introduce operational limit violations, primarily driven by voltage magnitude constraints. The implemented flexibility strategy effectively mitigates these violations through coordinated active and reactive power control, enhancing system security and operational efficiency. The findings highlight the potential of DT-based approaches for reliable and flexible operation of future distribution networks.
\end{abstract}


\section{Introduction}

Power distribution systems are undergoing a profound transformation driven by the increasing penetration of DERs, such as photovoltaic generation, wind power, and flexible demand. This transition is reshaping traditionally passive networks into actively managed systems characterized by reduced inertia, bidirectional power flows, and highly variable operating conditions \cite{Javed2021DERChallenges}. As a result, distribution system operators face increasing challenges in maintaining voltage limits and ensuring secure operation under dynamic and uncertain operating conditions \cite{Sun2019VoltageControl}.
Traditional “fit-and-forget” approaches based on network reinforcement and offline planning lack the responsiveness required to manage these dynamics in real time. Consequently, advanced operational tools are needed to continuously monitor network conditions and coordinate distributed resources for secure and reliable operation \cite{Xiang2016OptimalADN}, \cite{Karagiannopoulos2017HybridADN}.
Digital Twins have emerged as a promising solution for addressing these challenges. The Digital Twin (DT) is a virtual representation of a physical system that is continuously synchronized with real-world measurements, enabling enhanced situational awareness and decision support \cite{Zomerdijk2023DTArchitecture}, \cite{Thwe2025DTReview}. By integrating measurement data with network models, DTs support applications such as system monitoring and operational planning \cite{Song2023DTOverview}. Despite this potential, there are limited existing DT applications in distribution systems that support advanced operational decision-making. 
In particular, challenges persist in integrating real-time measurements with network-constrained analysis \cite{Yan2024DTImplementation}, \cite{Bragatto2023NearRealTimeDT}.
To address these gaps, this paper implements a real-time DT approach for active distribution networks, integrating measurement data with a network model to enable security assessment and coordinated corrective and preventive flexibility activation under network constraints. The implemented approach provides a unified operational view linking system monitoring, contingency analysis, and flexibility-based control actions. Its applicability is demonstrated on the Bornholm distribution system using real measurement data.
The remainder of this paper is organized as follows. Section II presents the implemented methodology. Section III describes the Bornholm distribution grid case study. Section IV outlines the study scenarios. Section V presents and discusses the results. Finally, Section VI concludes the paper.

\section{Methodology}

\subsection{Digital Twin Model of the Distribution Network}

DTs enable the representation and analysis of physical power systems through continuously updated digital counterparts driven by operational measurements. In this work, a DT of the Bornholm distribution network is developed using detailed network data provided by the Distribution System Operator (DSO) TREFOR and real-time measurements from smart metering infrastructure delivered by TREFOR and supported by the local utility company BEOF.
The dataset includes the electrical parameters and topology of the 60~kV distribution network, covering the external interconnection, buses, lines and cables, 60/10~kV transformers, and shunt elements. Based on this information, the complete network model is reconstructed in \texttt{pandapower}, an open-source Python-based power system analysis framework \cite{pandapower2018}. The framework provides component models and built-in routines for AC power flow and contingency analysis, enabling flexible and reproducible DT implementation.
The modeled network explicitly represents the system up to the 60/10 kV substations. Downstream medium-voltage feeders connected to the 10 kV side are not modeled individually; instead, their aggregated electrical behavior is represented by equivalent load and generation units at the corresponding buses. This aggregation preserves the net power exchange at each substation interface while maintaining a computationally efficient representation suitable for system-level security and flexibility analysis.

\subsection{Data Integration and Digital Twin Architecture}

The DT is continuously updated using real-time measurements obtained from the EnergyDataDK (EDDK) platform. EDDK is an open data platform for collecting, managing, and sharing energy-related measurements \cite{eddk}. It serves as the data backbone of the DT architecture and is part of the PowerLabDK infrastructure at the Technical University of Denmark (DTU), enabling integration of real-time measurements from the Bornholm distribution network via API and MQTT-based interfaces. 
Through EDDK, active power measurements from individual 10 kV feeders are obtained from the smart metering infrastructure with a temporal resolution of 15 minutes. For each feeder, two time-series are provided: injected power (generation) and withdrawn power (consumption). These measurements are aggregated at each 60/10 kV substation and used to update the active power injections of the equivalent load and generation units in the DT. 
Since reactive power measurements are not available, reactive power injections are reconstructed using fixed power factor assumptions. For loads, a lagging power factor of $0.95$ is assumed to represent the typical composition of residential and small-scale industrial demand in the MV network. For generation, a power factor of $0.99$ is used to capture the aggregated behavior of downstream DERs while limiting excessive reactive power injection at the substation interface and preserving sufficient flexibility margins for voltage control.

\begin{figure}[!ht]
\centering
\includegraphics[width=3.5in]{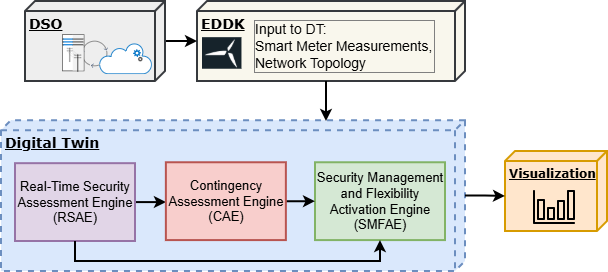}
\caption{Integration of the developed digital twin with the EnergyDataDK (EDDK) as a data space.}
\label{DT}
\end{figure} 

The DT architecture is organized around three functional engines:

\begin{itemize}
    \item Real-Time Security Assessment Engine (RSAE)
    \item Contingency Assessment Engine (CAE)
    \item Security Management and Flexibility Activation Engine (SMFAE)
\end{itemize}

The functionality and implementation of these engines are described in the following subsections. Their interaction with the EDDK data infrastructure is illustrated in Fig.~\ref{DT}.

\subsection{Real-Time Security Assessment Engine (RSAE)}

The RSAE constitutes the operational core of the DT and continuously evaluates the security state of the Bornholm distribution network. It operates on the \texttt{pandapower}-based network model whose operating point is updated using the active power measurements and reconstructed reactive power values described in the previous subsection. The operating point at each timestamp is defined by these active power measurements and reconstructed reactive power injections used to update the network model. Using this updated state, the RSAE performs an AC power-flow calculation with the Newton–Raphson method provided by \texttt{pandapower}, yielding bus voltage magnitudes and angles, branch power flows, and transformer loading levels.
These results are compared against predefined operational security limits: bus voltages within $0.95$–$1.05$ p.u. and thermal loading limits of $90\%$ of rated capacity for lines and transformers. Any deviations beyond these limits are classified as violations and forwarded to the SMFAE for corrective flexibility activation, as illustrated in Fig.~\ref{DT}.

\subsection{Contingency Assessment Engine (CAE)}

The CAE assesses network security under potential contingency scenarios using the RSAE operating point as the baseline. It systematically analyzes $N\!-\!1$ outages, corresponding to the loss of a single line or transformer. For each contingency, the affected network topology is updated, and an AC power-flow analysis is performed in \texttt{pandapower}. The resulting state is evaluated against the same operational security limits used in the RSAE to ensure consistency between normal and contingency assessments. Any deviations beyond these limits are classified as violations and forwarded to the SMFAE for preventive flexibility activation.
Some contingencies may result in islanding of portions of the distribution network. The CAE is therefore designed to identify vulnerable operating conditions and quantify the operational impact of credible outages, rather than enforcing strict $N\!-\!1$ security.

\subsection{Security Management and Flexibility Activation Engine (SMFAE)}

The SMFAE is the decision-making layer of the DT, ensuring secure and reliable network operation by processing violations identified by the RSAE and CAE. It uses a common optimization formulation, applied separately to each case. The model computes optimal active and reactive power setpoints ($P$ and $Q$) for generators at each substation. These setpoints are determined to resolve the identified violations through coordinated active and reactive power adjustments, enabling corrective and preventive flexibility activation. If no feasible solution exists, the problem is declared infeasible. The detailed mathematical formulation of the optimization problem is provided in the next subsection.

\subsection{AC Power Flow–Based Optimization Formulation}

The SMFAE employs an AC power flow–based optimization formulation to determine the redispatched active and reactive power setpoints ($P$ and $Q$) of generators. The objective is to resolve operational limit violations while minimizing the overall redispatch effort. The formulation captures the nonlinear relationships between nodal power injections, voltage magnitudes and angles, and branch flows. \newline

\subsubsection*{Sets, Parameters, and Decision Variables}

The optimization problem is defined using the following sets, parameters, and decision variables:

\begin{itemize}

\item $\mathcal{N}$: set of all buses

\item $\mathcal{E}$: set of network branches (lines and transformers) connecting buses $(i,j) \in \mathcal{N}$

\item $\mathcal{G}$: set of generators

\item $\mathcal{L}$: set of loads

\item $\mathcal{G}_i \subseteq \mathcal{G}$: set of generators connected to bus $i \in \mathcal{N}$

\item $\mathcal{L}_i \subseteq \mathcal{L}$: set of loads connected to bus $i \in \mathcal{N}$

\item $G_{ij}, B_{ij}$: real and imaginary components of the bus admittance matrix $Y_{ij}=G_{ij}+jB_{ij}$

\item $|Y_{ij}|$: magnitude of the series admittance of branch $(i,j) \in \mathcal{E}$

\item $I_{ij}^{\max}$: thermal current limit of branch $(i,j) \in \mathcal{E}$

\item $P_l^{L}, Q_l^{L}$: active and reactive power demand of load $l \in \mathcal{L}$

\item $P_g^{\text{base}}, Q_g^{\text{base}}$: base (pre-redispatch) active and reactive power output of generator $g \in \mathcal{G}$

\item $P_g^{\text{hist,max}}$: historical maximum active power output of generator $g \in \mathcal{G}$

\item $w_P, w_Q$: weighting coefficients for active and reactive power redispatch

\item $P_i^{\text{ext}}, Q_i^{\text{ext}}$: net active and reactive power exchanged with the external grid at the slack bus $i = 0 \in \mathcal{N}$

\item $Q_i^{\text{sh}}$: net reactive power consumption of shunt elements at bus $i \in \mathcal{N}$ (positive for inductive behavior, negative for capacitive injection)

\item $P_g^{\text{new}}, Q_g^{\text{new}}$: redispatched active and reactive power output of generator $g$ (decision variables)

\item $V_i, \theta_i$: voltage magnitude and phase angle at bus $i \in \mathcal{N}$ (decision variables)

\end{itemize}

\subsubsection*{Objective Function}

The redispatch effort is modeled through the following quadratic cost function:

\begin{align}
\min \sum_{g \in \mathcal{G}} 
\left[ 
w_P \left( P_g^{\text{new}} - P_g^{\text{base}} \right)^2 
+ 
w_Q \left( Q_g^{\text{new}} - Q_g^{\text{base}} \right)^2 
\right]
\end{align}

Active power adjustments are penalized more strongly due to their direct impact on line loading, while reactive power redispatch provides flexibility for voltage regulation. \newline

\subsubsection*{Constraints}

The optimization is subject to the following nonlinear constraints:

\iffalse
\paragraph{Active Power Balance:}
\begin{equation}
\begin{split}
\sum_{g \in \mathcal{G}_i} P_g^{\text{new}} 
+ P^{\text{ext}} 
- \sum_{l \in \mathcal{L}_i} P_l^{L} 
&= \sum_{j \in \mathcal{N}} V_i V_j 
\big[
G_{ij} \cos(\theta_i - \theta_j) \\
&\quad + B_{ij} \sin(\theta_i - \theta_j)
\big],
\quad \forall i \in \mathcal{N}
\end{split}
\end{equation}
\fi

\paragraph{Active Power Balance}
\begin{multline}
\sum_{g \in \mathcal{G}_i} P_g^{\text{new}} 
+ P_i^{\text{ext}} 
- \sum_{l \in \mathcal{L}_i} P_l^{L} 
= \sum_{j \in \mathcal{N}} V_i V_j 
\Big[
G_{ij} \cos(\theta_i - \theta_j) \\
+ B_{ij} \sin(\theta_i - \theta_j)
\Big], \forall i \in \mathcal{N}
\end{multline}

\iffalse
\paragraph{Reactive Power Balance}
\begin{equation}
\begin{split}
\sum_{g \in \mathcal{G}_i} Q_g^{\text{new}} 
+ Q^{\text{ext}} 
- \sum_{l \in \mathcal{L}_i} Q_l^{L}
- Q_i^{\text{sh}}
&= \sum_{j \in \mathcal{N}} V_i V_j
\big[
G_{ij} \sin(\theta_i - \theta_j) \\
& \quad - B_{ij} \cos(\theta_i - \theta_j)
\big],
\quad \forall i \in \mathcal{N} 
\end{split}
\end{equation}
\fi

\paragraph{Reactive Power Balance}
\begin{multline}
\sum_{g \in \mathcal{G}_i} Q_g^{\text{new}} + Q_i^{\text{ext}} - \sum_{l \in \mathcal{L}_i} Q_l^{L} - Q_i^{\text{sh}} = 
\sum_{j \in \mathcal{N}} V_i V_j \Big[ G_{ij} \sin(\theta_i - \\ \theta_j) - B_{ij} \cos(\theta_i - \theta_j) \Big], \forall i \in \mathcal{N}
\end{multline}

\paragraph{Branch Flow Limits}

To ensure secure operation and maintain a safety margin, branch currents are limited to 90\% of their thermal ratings.

\iffalse
\begin{equation}
|Y_{ij}|^2 \left( V_i^2 + V_j^2 - 2 V_i V_j \cos(\theta_i - \theta_j) \right)
\leq (0.9 I_{ij}^{\text{max}})^2, 
\quad \forall (i,j) \in \mathcal{E}
\end{equation}
\fi

\begin{multline}
|Y_{ij}|^2 \Big( V_i^2 + V_j^2 
- 2 V_i V_j \cos(\theta_i - \theta_j) \Big)
\leq (0.9\, I_{ij}^{\text{max}})^2, \\ \forall (i,j) \in \mathcal{E}
\end{multline}

\paragraph{Generator Operating Limits}

Active power redispatch of generators is bounded within $\pm 85\%$ of their historical maximum to ensure feasible operation while preserving flexibility for corrective and preventive actions:

\begin{align}
-0.85\, P_g^{\text{hist,max}} \leq P_g^{\text{new}} \leq 0.85\, P_g^{\text{hist,max}}, \quad \forall g \in \mathcal{G}
\end{align}

Reactive power capability is coupled to active power through a power-factor–based constraint reflecting inverter apparent power limits. The feasible operating region is defined by a minimum power factor of $|PF| = 0.95$:

\begin{align}
Q_g^{\text{new}} \leq \phantom{-}\tan\!\left(\arccos(0.95)\right)\, \lvert P_g^{\text{new}} \rvert, \quad \forall g \in \mathcal{G} \\
Q_g^{\text{new}} \geq -\tan\!\left(\arccos(0.95)\right)\, \lvert P_g^{\text{new}} \rvert, \quad \forall g \in \mathcal{G}
\end{align}

\paragraph{Voltage Limits}
\begin{align}
0.95 \leq V_i \leq 1.05, \quad \forall i \in \mathcal{N} \\
-\pi \leq \theta_i \leq \pi, \quad \forall i \in \mathcal{N} \\
\theta_{i=0} = 0, \quad \forall i \in \mathcal{N}
\end{align}

\subsubsection*{Implementation}
The model is implemented in \texttt{Pyomo} and solved using \texttt{IPOPT} \cite{pyomo,ipopt}. The use of a nonlinear interior-point solver enables efficient solution of the AC power flow–based optimization problem with continuous decision variables. This formulation ensures secure operation under both normal and contingency conditions while maintaining computational tractability.

\section{Case Study: Bornholm Distribution Grid}

Bornholm is a Danish island located in the Baltic Sea and is widely used as a test system for renewable energy integration and smart grid operation. The island has approximately 40,000 inhabitants, an annual electricity demand of about 250~GWh, and a peak load of approximately 55~MW. Electricity is supplied to roughly 28,000 customers through a distribution network comprising 16 primary substations \cite{PowerLabDK2022SmartCommunityBornholm}.

The island is interconnected with the Swedish transmission system via a 60~MW, 60~kV subsea HVAC cable connecting Bornholm to the 135~kV grid at Borrby in southern Sweden over a distance of approximately 50~km. This interconnection enables both grid-connected and islanded operation, making the system suitable for studying security assessment and flexibility management.

Internally, the network is structured around a 60~kV backbone ring interconnecting the 16 primary substations, each stepping down to 10~kV. The downstream medium- and low-voltage networks supply residential, commercial, and industrial consumers and include distributed energy resources such as wind, photovoltaic, and biogas-based generation, as well as dispatchable units (e.g., CHP and backup generators) that provide operational flexibility \cite{PowerLabDK2022SmartCommunityBornholm, Gabderakhmanova2020Bornholm}.

%Figure~\ref{BH_ref} illustrates the geographical distribution of substations and representative generation units.

In this study, the developed Digital Twin (DT) model is applied to the higher-voltage (60 kV) distribution network, with downstream feeders represented by equivalent load and generation at 10~kV level, as described in Section II.

\section{Study Scenarios}

The performance of the implemented DT is evaluated over one year of operating points derived from measurement-driven active power time series with a 15-minute temporal resolution, as discussed in Section II. The following study scenarios are defined to assess system behavior under reference operating conditions, demand uncertainty, and contingency conditions.

\subsection{Base Case}

A Base Case is established by performing AC power flow simulations over the study horizon and applying minor adjustments where necessary to ensure violation-free operation. This case represents a secure reference operating point of the network and serves as the input for contingency analysis under $N\!-\!1$ outage scenarios.

\subsection{Load Variation Scenarios}

To assess system behavior under demand uncertainty, two additional scenarios are defined by uniformly scaling the Base Case load profile by $+20\%$ and $-20\%$. These variations are applied across the study horizon while preserving the network topology and operational assumptions.

\begin{figure}[!ht]
\centering
\includegraphics[width=3.5in, height=1.7in]{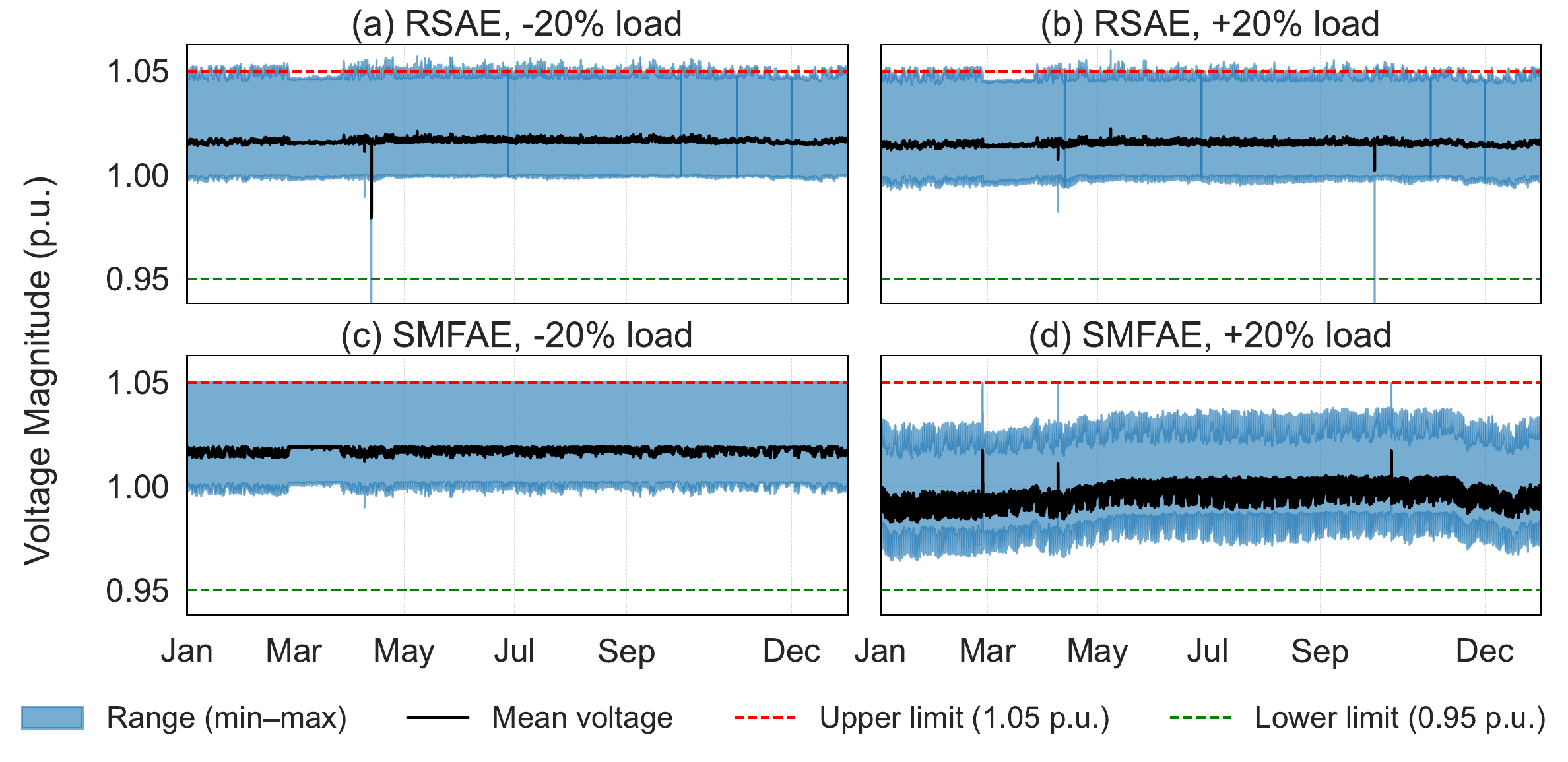}
\caption{Voltage magnitude envelope at network busbars under the ±20\% load variation scenarios: (a)–(b) RSAE and (c)–(d) SMFAE.}
\label{volt_rsae}
\end{figure}

\section{Results and Discussion}

The results are organized according to the three core engines of the DT to reflect its functional workflow. First, the Real-Time Security Assessment Engine (RSAE) evaluates the system under the Load Variation Scenarios, identifying violations caused by demand uncertainty. Next, the Contingency Assessment Engine (CAE) assesses system security under $N\!-\!1$ line and transformer outages using the Base Case as the operating point. Finally, the Security Management and Flexibility Activation Engine (SMFAE) determines optimal corrective and preventive actions to mitigate the identified violations by RSAE and CAE to restore secure operation.
This structure emphasizes the sequential interaction among the three engines, from violation detection to flexibility activation, and highlights the role of coordinated flexibility management in maintaining secure network operation.

\begin{figure}[!ht]
\centering
\includegraphics[width=3.5in]{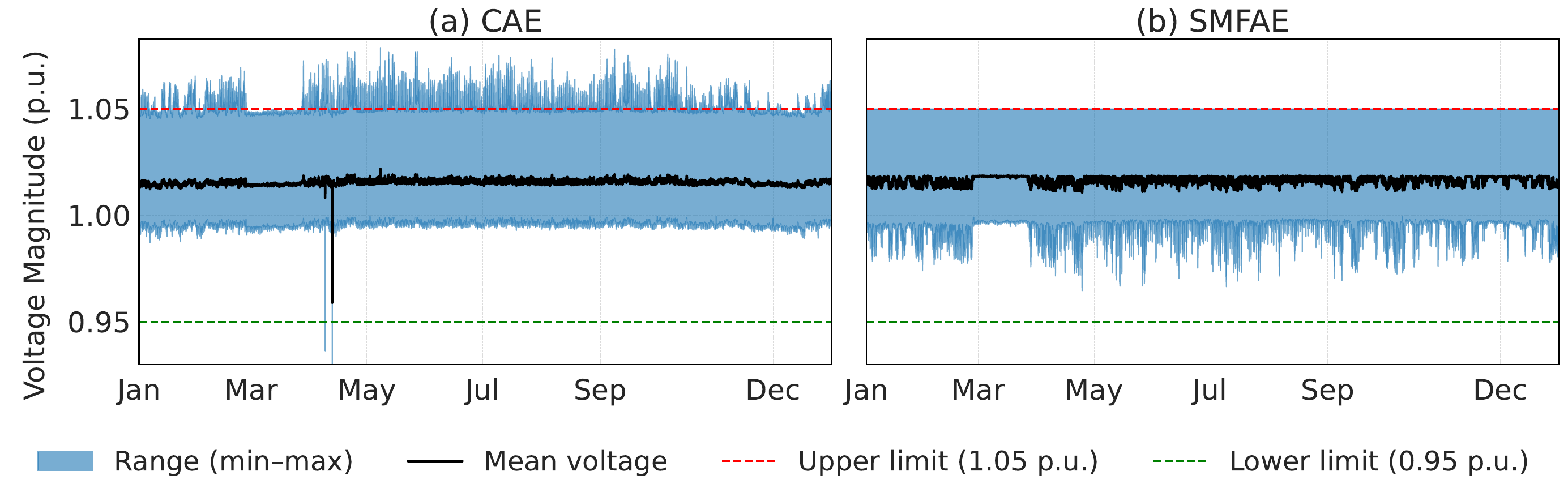}
\caption{Voltage magnitude envelope at network busbars under $N\!-\!1$ contingency conditions: (a) CAE and (b) SMFAE.}
\label{volt_ca}
\end{figure}

\subsection{Results of RSAE}

The RSAE is evaluated under the Load Variation Scenarios to assess voltage and thermal limit violations. No thermal violations are observed; therefore, the analysis focuses on voltage magnitudes outside the permissible range of 0.95--1.05~p.u. As shown in Fig.~\ref{volt_rsae}(a), the $-20\%$ load variation scenario results in overvoltage violations in 1.69\% of the operating points, with values ranging from 1.05 to 1.06~p.u. For the $+20\%$ load variation scenario in Fig.~\ref{volt_rsae}(b), overvoltage violations occur in 0.68\% of the operating points, with values between 1.05 and 1.06~p.u. These violations are subsequently provided to the SMFAE for corrective flexibility activation.

\subsection{Results of CAE}

The CAE is evaluated using the Base Case under $N\!-\!1$ contingency conditions involving single line and transformer outages, to identify operational violations related to voltage and thermal limits. A total of 39 contingencies (23 line and 16 transformer outages) are analyzed based on the network topology. No thermal violations are observed, and the results indicate that violations are predominantly associated with bus voltage magnitudes, as shown in Fig.~\ref{volt_ca}(a). Overvoltage violations occur in 1.42\% of the operating points, with values ranging from 1.05 to 1.08~p.u., while undervoltage occurrences are negligible. These violations are addressed by the SMFAE through preventive flexibility activation.

\subsection{Results of SMFAE}

\begin{figure}[!ht]
\centering
\includegraphics[width=3.5in, height=2.3in]{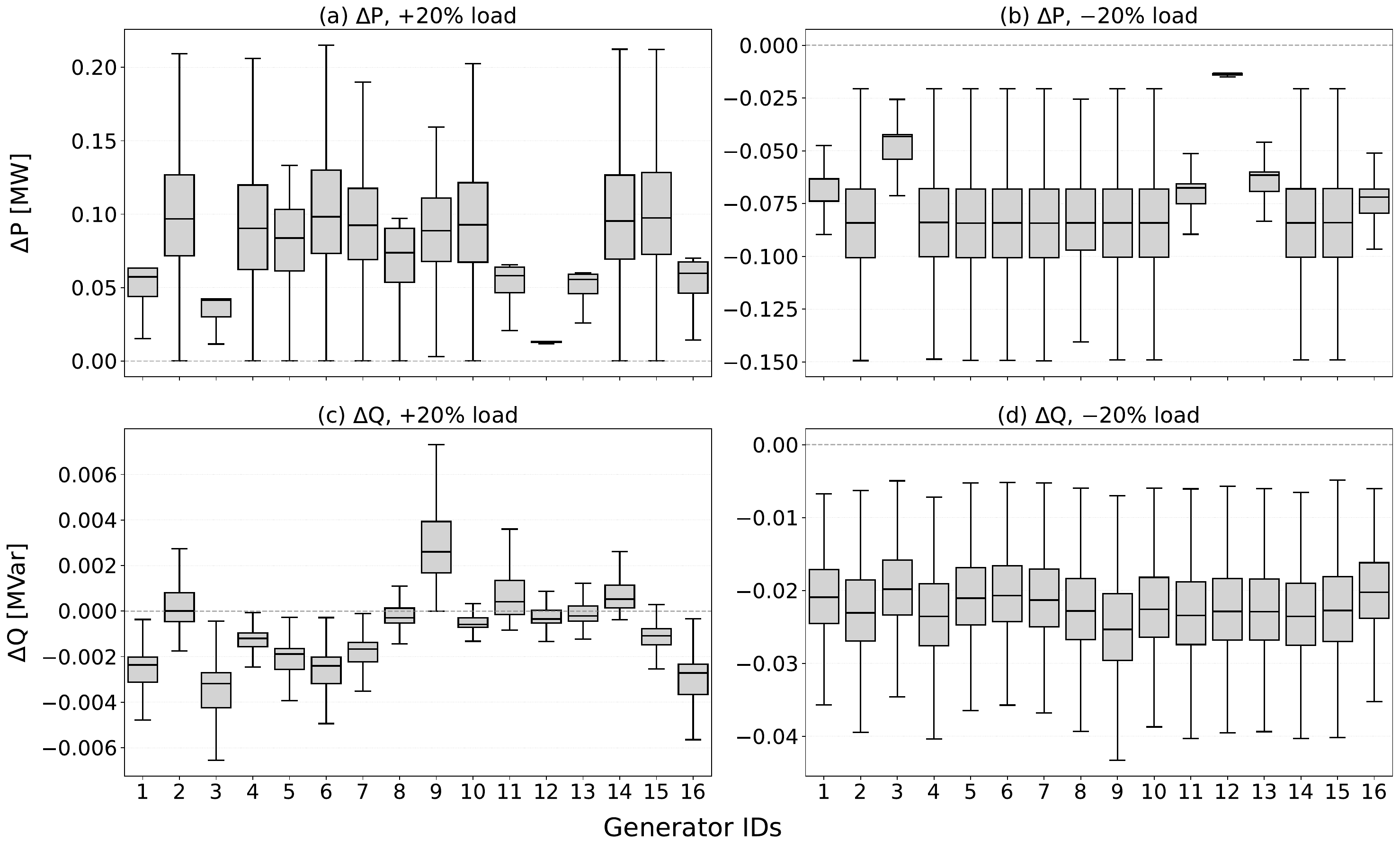}
%\caption{Distribution of active and reactive power adjustments across substations under the $\pm20\%$ load variation scenarios: (a) $+20\%$ and (b) $-20\%$, obtained using the SMFAE.}
\caption{
Distribution of active and reactive power adjustments across generators under $\pm20\%$ load variation scenarios using SMFAE:
(a) $\Delta P$ for $+20\%$ load,
(b) $\Delta P$ for $-20\%$ load,
(c) $\Delta Q$ for $+20\%$ load,
(d) $\Delta Q$ for $-20\%$ load.
}
\label{flex_deviation_kwh}
\end{figure}

The SMFAE is evaluated based on the violations identified by the RSAE under the Load Variation Scenarios and by the CAE, which is evaluated using the Base Case under $N\!-\!1$ contingency conditions. In all cases, the observed violations are predominantly associated with overvoltage conditions.
To mitigate these violations, the SMFAE determines optimal active ($P$) and reactive ($Q$) power setpoints for the generators using the optimization model described in Section II. The objective is to compute new setpoints such that the constraints defined in the optimization model are satisfied while maintaining the external grid import at the Base Case level, with the required balancing achieved through coordinated adjustments in local generation.
Fig.~\ref{flex_deviation_kwh} presents the distribution of $\Delta P = P_g^{new} - P_g^{base}$ and $\Delta Q = Q_g^{new} - Q_g^{base}$ over the study horizon for all generators across the 16 substations. Here, $P_g^{new}$ and $Q_g^{new}$ denote the optimal setpoints obtained by SMFAE to resolve the violations identified by the RSAE under Load Variation Scenarios, while $P_g^{base}$ and $Q_g^{base}$ represent the corresponding Base Case setpoints. In Fig.~\ref{flex_deviation_kwh}, (a) and (c) correspond to the $+20\%$ load variation scenario, whereas (b) and (d) represent the $-20\%$ load variation scenario. Overall, the results show that the optimized dispatch redistributes generation across the network to mitigate voltage violations and maintain secure operation.

\begin{figure}[!ht]
\centering
\includegraphics[width=3.5in]{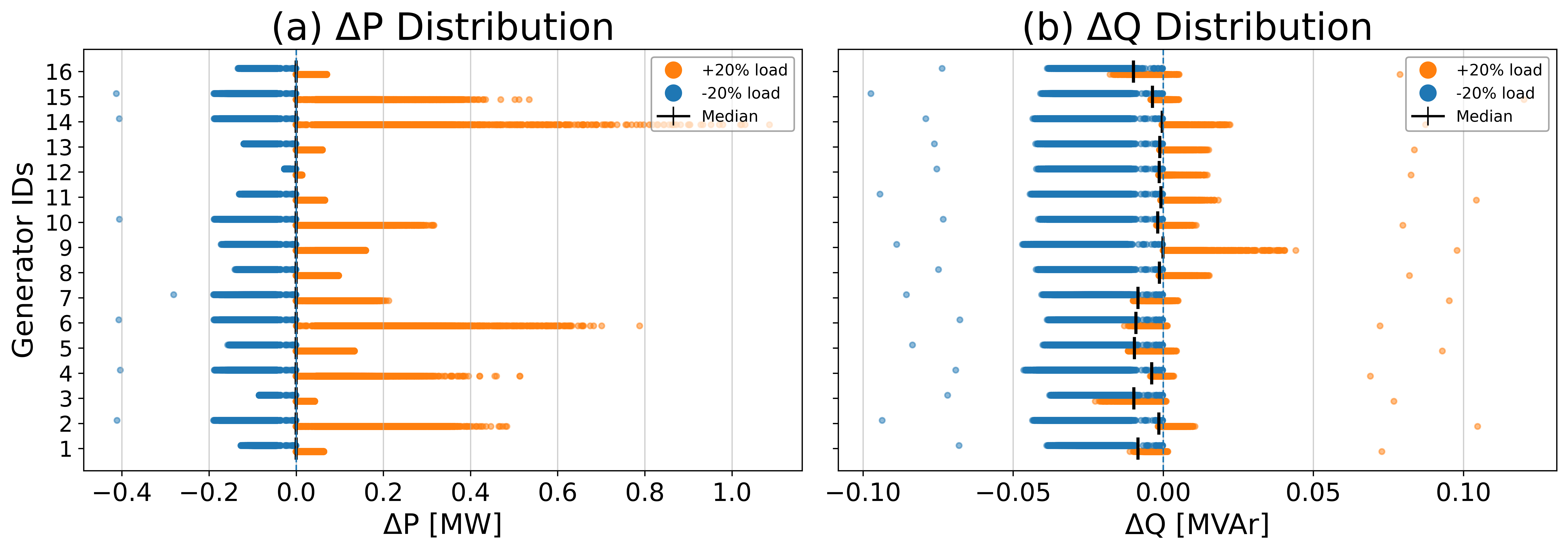}
\caption{Frequency and direction of active and reactive power distribution across generators under $\pm20\%$ load variation scenarios using SMFAE: (a) $\Delta P$, (b) $\Delta Q$.}
\label{delta_PQ_com}
\end{figure}

The boxplots reveal distinct trends across demand scenarios. For the $+20\%$ load variation scenario, $\Delta P$ is positive for all generators, reflecting increased active power generation to meet higher demand. In contrast, $\Delta Q$ exhibits both positive and negative values, indicating that reactive power adjustments are not uniformly correlated with load increase. This behavior reflects the voltage-sensitive characteristics of the Bornholm network and highlights the non-trivial role of reactive power in voltage regulation. 
For the $-20\%$ load variation scenario, both $\Delta P$ and $\Delta Q$ are predominantly negative, consistent with reduced generation under lower demand conditions. Across both $\pm20\%$ load variation scenarios, the distribution of $\Delta P$ at the individual generator level indicates heterogeneous participation in flexibility activation. For instance, Generator~12 shows minimal and nearly constant contribution across the year, as indicated by its narrow boxplot, suggesting a limited role in mitigating network violations. In contrast, generators such as 2, 6, and 10 exhibit larger and more variable contributions, indicating their greater importance in providing flexibility and adapting to temporal variations in system conditions.
The results provide insight into the distribution and direction of flexibility activation derived from the SMFAE. The analysis considers only non-zero activations, i.e., time steps where flexibility is effectively utilized. Fig.~\ref{delta_PQ_com} shows the empirical distribution of $\Delta P$ and $\Delta Q$ values for each generator under the $+20\%$ and $-20\%$ load variation scenarios, with median values computed over the full annual dataset.
Fig.~\ref{delta_PQ_com}a shows $\Delta P$ is consistently positive under the $+20\%$ load variation scenario and negative under the $-20\%$ load variation scenario, resulting in median values close to zero across all generators. This indicates a symmetric and balanced deployment of active power flexibility. In contrast, Fig.~\ref{delta_PQ_com}b shows that reactive power exhibits both positive and negative $\Delta Q$ values under both load variation scenarios, with median values tending toward the negative range. This behavior reflects the inherently asymmetric nature of reactive power regulation, which is primarily driven by voltage control requirements rather than directly by load variation.

\begin{figure}[!ht]
\centering
\includegraphics[width=3.5in, height=0.8in]{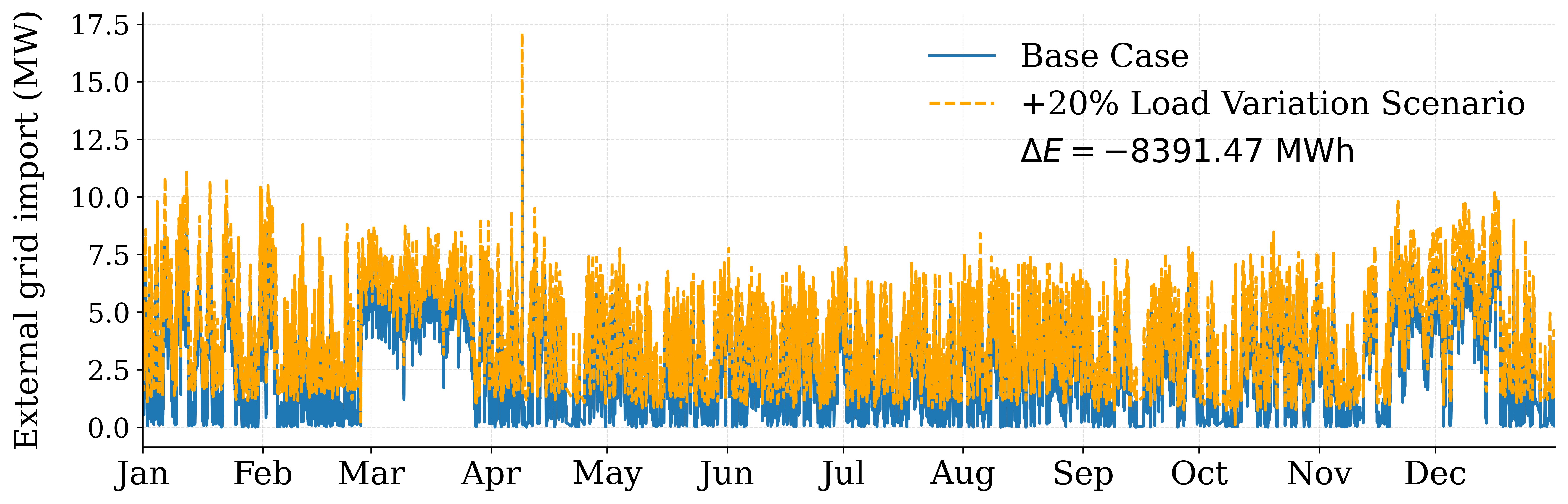}
\caption{Comparison of external grid import between the Base Case and the +20\% load variation scenario over the study period, prior to flexibility activation.}
\label{slack_power}
\end{figure}

Under the $\pm20\%$ load variation scenarios, the application of optimal setpoints by SMFAE eliminates the overvoltage violations identified by the RSAE, as shown in Fig.~\ref{volt_rsae}(c) and (d), where all bus voltages remain within the permissible range over the study horizon. In this context, the resulting setpoints also provide an estimate of day-ahead flexibility requirements under varying demand conditions.
For the $+20\%$ load variation scenario, the optimized dispatch reduces reliance on external grid support, achieving a total annual reduction of 8391.47~MWh in energy imports compared to operation without SMFAE activation. As illustrated in Fig.~\ref{slack_power}, external grid import would otherwise increase significantly under higher demand conditions in the absence of flexibility activation. This demonstrates the effectiveness of coordinated active and reactive power control in enhancing both operational security and energy efficiency. Similarly, under the $-20\%$ load variation scenario, the external grid import remains at the Base Case level, while local generators reduce their active and reactive outputs accordingly.

\begin{figure}[!ht]
\centering
\includegraphics[width=3.5in, height=1.2in]{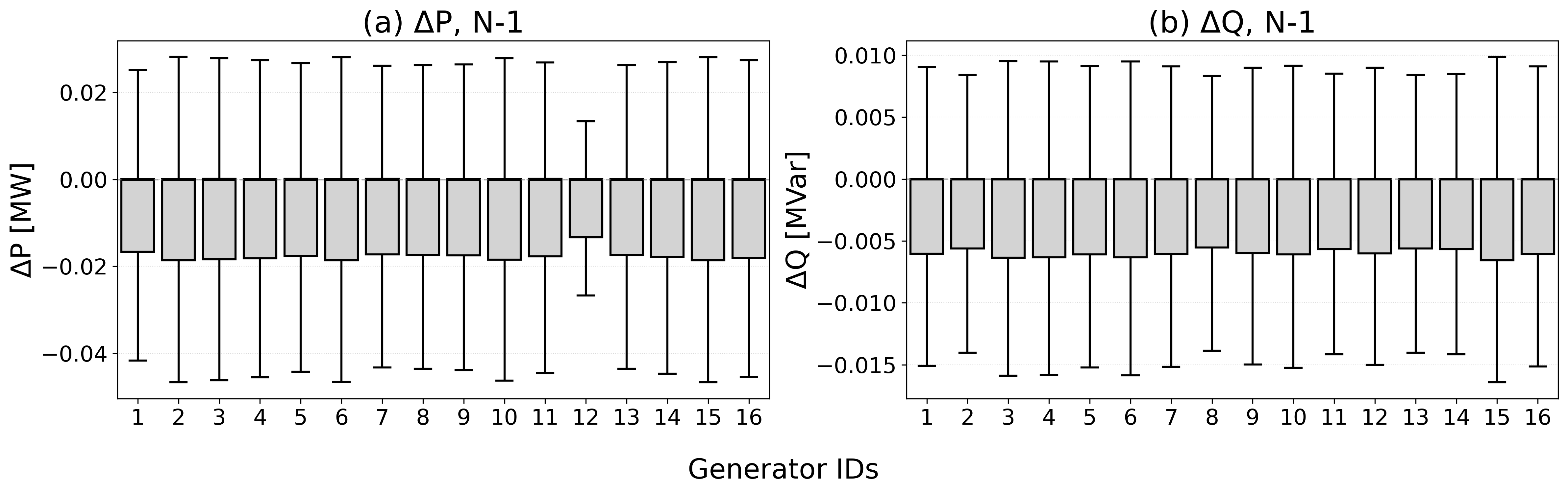}
\caption{Distribution of active and reactive power adjustments across generators under $N\!-\!1$ contingency conditions using SMFAE.}
\label{n_1_flex_res}
\end{figure}

Fig.~\ref{n_1_flex_res} presents the distribution of $\Delta P$ and $\Delta Q$ for CAE-based mitigation using SMFAE. The optimal setpoints resolve violations under $N\!-\!1$ contingency conditions using the Base Case as reference. As shown in Fig.~\ref{volt_ca}(b), the resulting dispatch eliminates overvoltage violations and maintains all bus voltages within limits.
Under contingency conditions, the SMFAE maintains the external grid import at the Base Case level and compensates for component outages through coordinated adjustments of active and reactive power using local flexibility resources. This ensures that contingency-induced imbalances are managed without increasing reliance on external grid support.
The boxplots show that the mean values of $\Delta P$ are predominantly close to zero across generators. This behavior is linked to the network structure, where equivalent generation and load are connected at the 10~kV level of each 60/10~kV substation. Consequently, outages of lines or transformers often result in the simultaneous loss of both generation and load at affected substations. The remaining demand is redistributed among the generators that remain connected, leading to an overall reduction in active power output relative to the Base Case. A similar trend is observed for $\Delta Q$, indicating coordinated reactive power adjustments to maintain voltage security.
Overall, the results demonstrate that the SMFAE effectively manages contingency-induced operational stress through distributed flexibility while preserving the external grid exchange at its nominal level.

\section{Conclusion}

This paper presented a Digital Twin (DT) approach, developed using open-source tools, for real-time security assessment and corrective and preventive flexibility activation in the Bornholm distribution network using one year of smart meter data with 15-minute resolution. A secure Base Case is established through minor adjustments and used as the reference operating point. The RSAE results show limited overvoltage violations affecting 1.69\% and 0.68\% of operating points for the $-20\%$ and $+20\%$ load variation scenarios, respectively, with voltages up to 1.06~p.u. Under $N\!-\!1$ contingencies, the CAE identifies overvoltage violations in 1.42\% of operating points, reaching 1.08~p.u., while thermal limits remain satisfied. The SMFAE mitigates these violations through coordinated active and reactive power adjustments while maintaining the external grid exchange at the Base Case level. Under the $+20\%$ load scenario, a non-uniform reactive power response is observed, reflecting the voltage-sensitive behavior of the Bornholm network. The implemented approach also enables estimation of day-ahead flexibility requirements under $\pm20\%$ load variations, supporting preventive operation, reducing reliance on external grid support, and improving operational efficiency. These capabilities are achieved within computational times compatible with real-time operation, remaining below 140 seconds. Overall, the implemented DT demonstrates the effectiveness of data-driven flexibility activation for enhancing voltage security and supporting operational decision-making in active distribution networks.

\iffalse
\section*{Acknowledgment}

This work was carried out as part of the SYNERGIES project, funded by the European Union’s Horizon Europe programme under Grant Agreement No. 101069839, and by the ODEON project, funded by the European Union's LCE Policy Support Programme under Grant Agreement No. 101136128.
\fi

\section*{Acknowledgment}
The authors acknowledge the support of the SYNERGIES and ODEON projects.

\iffalse

\fi

\end{document}